\documentclass[9pt,twocolumn,twoside]{osajnl}

\journal{ol} 

\setboolean{shortarticle}{true}

\title{Neural Computing with Coherent Laser Networks}

\author[1,2,*]{Mohammad-Ali Miri}
\author[3,2]{Vinod Menon}

\affil[1]{Department of Physics, Queens College of the City University of New York, Queens, New York 11367, USA}%
\affil[2]{Physics Program, The Graduate Center, City University of New York, New York, New York 10016, USA}
\affil[3]{Department of Physics, City College of the City University of New York, New York, NY 10031, USA}
\affil[*]{Corresponding author: mmiri@qc.cuny.edu}

\affil[*]{mmiri@qc.cuny.edu}

\begin{abstract}

We show that a coherent network of lasers exhibits emergent neural computing capabilities. The proposed scheme is built on harnessing the collective behavior of laser networks for storing a number of phase patterns as stable fixed points of the governing dynamical equations and retrieving such patterns through proper excitation conditions, thus exhibiting an associative memory property. The associative memory functionality is first discussed in the strong pumping regime of a network of passive dissipatively coupled lasers which simulate the classical XY model. It is discussed that despite the large storage capacity of the network, the large overlap between fixed-point patterns effectively limits pattern retrieval to only two images. Next, we show that this restriction can be uplifted by using nonreciprocal coupling between lasers and this allows for utilizing a large storage capacity. This work opens new possibilities for neural computation with coherent laser networks as novel analog processors. In addition, the underlying dynamical model discussed here suggests a novel energy-based recurrent neural network that handles continuous data as opposed to Hopfield networks and Boltzmann machines that are intrinsically binary systems.

\end{abstract}

\begin{document}

\maketitle

\section{Introduction}

In the recent years, there has been a growing interest in developing new platforms for general-purpose or application-specific computing that offer an advantage over classical processors in terms of computational time, energy efficiency and scalability \cite{calude2017unconventional}. Although quantum computing is widely considered as a promising route, it appears that the classical nonlinear systems exhibit a largely under-explored computational capacity that is not properly utilized in conventional digital computers \cite{haken2004synergetic}. In this regard, there is great interest in developing alternative hardware platforms, which subsequently demand for compatible new algorithms.

Inspired by the biological brain, an interesting computational platform seems to be a network of nonlinear units, i.e., neurons, with a complex architecture that allows dense long-range interactions \cite{hertz2018introduction}. In such systems, computing is an emergent nonlinear dynamical behavior of the network, and, in principle, can be much more efficient for certain tasks in comparison with the well-established sequential architecture. Interestingly, in the physics community interest in the subject of neural computation was raised at an early stage by the introduction of Hopfield networks \cite{hopfield1982neural, hopfield1984neurons}. In these contexts, mainly influenced by spin systems in statistical mechanics, computing is viewed as finding states that minimize a global network energy function. Analog physical implementations of Hopfield networks with optoelectronics \cite{farhat1985optical} and CMOS circuits \cite{graf1987cmos, verleysen1989analog} were demonstrated for a small number of neurons at early stages. More importantly, such networks inspired unconventional methods for solving combinatorial optimization problems \cite{hopfield1985neural} as well as energy-based models for machine learning \cite{ackley1985learning}. On the other hand, interest in physical implementation of unconventional computing with densely connected architectures has recently regained interest in photonics \cite{marandi2014network, mcmahon2016fully, inagaki2016coherent}. In fact, energy-efficiency and the possibility of implementing long-range interactions make photonics an attractive candidate for neural computation. Accordingly, there is interest in developing novel methods and algorithms that allow for taking advantage of the existing photonics systems for unconventioanl computing.

\begin{figure*}[t]
    \centering
    \includegraphics[width=0.7\linewidth]{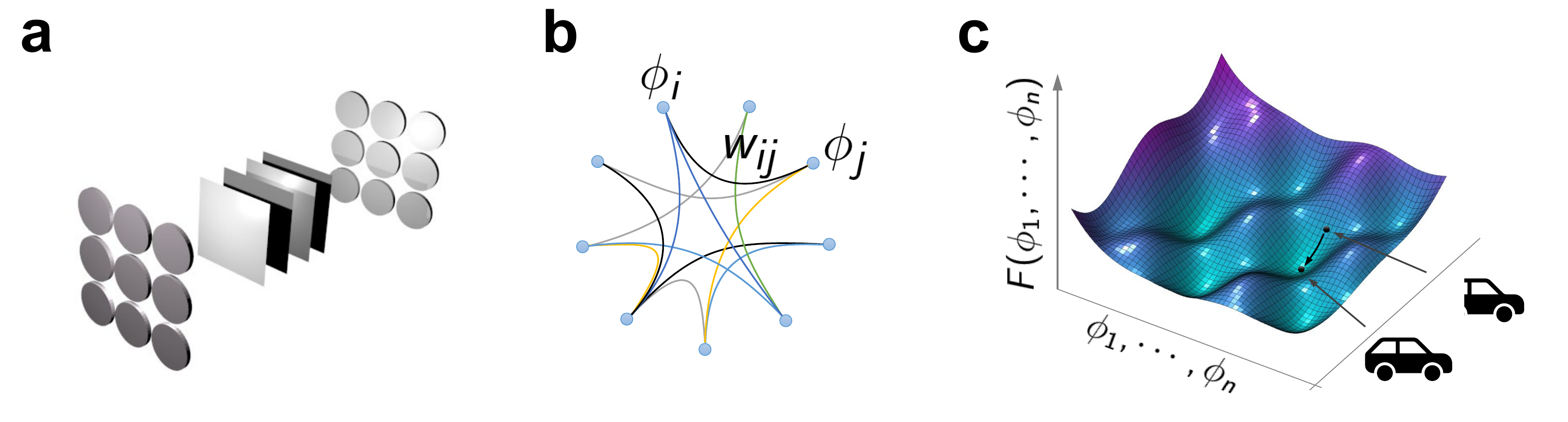}
    \caption{(a) A schematic of the coherent laser network, and (b) the associated network graph, and (c) the multivariate energy function. By locating a desired pattern at a local minimum of the energy function, it can be retrieved when the network is suitably initialized to start from the attractor basin of the embedded fixed point.}
    \label{fig1}
\end{figure*}

Here, we show that coherent laser networks exhibit collective neural computing capabilities, and devise the fundamental requirements for realizing an associative memory for continuous patterns. What facilitates this work is recent experimental progress in creating large networks of coherently coupled photonic oscillators \cite{marandi2014network, mcmahon2016fully, nixon2013observing, berloff2017realizing, parto2020realizing}. These activities have been primarily centered around solving computationally-hard problems by optical simulation of classical spin models. In particular, coherent laser networks have been used for solving non-convex optimization problems of the form of the classical XY Hamiltonian \cite{tradonsky2019rapid}, while numerical simulation of the governing dynamical models have been shown to be an efficient optimization method \cite{kalinin2018global}. Here, it is shown that coherent laser networks hold a great potential as a physical energy-based neural computing platform.

The present work is timely due to two important recent realizations that make coupled laser systems an attractive choice as a physical neural network. First, is the possibility of implementing dissipative interaction among laser networks which ensures the presence of fixed point attractors for such dynamical systems \cite{nixon2013observing, honari2020mapping}. The presence of dissipative coupling is shown to shift the dynamical model governing laser networks toward a class of reaction-diffusion systems that are known to be the host of exotic phenomena most notably pattern formation, which is the core of the present work \cite{honari2021self}. In contrast, driven by device applications, traditionally the general trend has been centered around dispersive interaction among laser arrays to avoid power loss, which in turn could result in unstable and chaotic behavior. Second, several recent works show the possibility of creating coupling through complex graph topologies, which is essential for implementing and training a recurrent neural network based on laser networks with desired wiring \cite{nixon2013observing, brunner2015reconfigurable}. In contrast, in the past the emphasis has been on lattice geometries with nearest neighbor couplings. It is because of this latter that we opt to call the system a \textit{laser network} rather than a \textit{laser array}.

Figure~\ref{fig1}(a) schematically depicts a network of $n$ lasers that are coherently coupled through diffraction engineering. This coherent laser network can be considered as a complex network represented with a graph as shown in Fig.~\ref{fig1}(b). Here, each graph node, represents an artificial neuron associated with a laser that is described by its amplitude and phase, $a_i(t) = |a_i(t)| \exp(i \phi_i (t))$, as two dynamical variables. In addition, two representative neurons $i$ and $j$, interact dynamically through rates $(w_{ij} , w_{ji})$, which could in general be non-reciprocal, i.e., $w_{ij} \neq w_{ji}$. Assuming that all lasers are identical, starting from an initial condition, under proper conditions the network can reach a phase-locking state where the amplitudes are nearly equal and the phases have a fixed pattern \cite{honari2020mapping}. In this regime, the system can be viewed as a network of phase oscillators that are governed by an $n$-dimensional energy landscape function as shown schematically in Fig.~\ref{fig1}(c). The equilibrium phase patterns of the laser network are associated with the local minima of this energy landscape function. Thus, the laser network can be viewed as an energy-based neural network. The use of such an energy-based model can be best demonstrated through an associative memory functionality. In such a system, by properly choosing the weight matrix, one can suitably engineer the landscape function such that desired patterns are located at its local minima as illustrated in Fig.~\ref{fig1}(d). In this manner, the network memorizes a given pattern which can be retrieved when it is suitably initialized.

In this work, first, it is shown that the conservative reciprocal coupling allows for the formation of binary patterns. We show that by using the Hebbian learning desired patterns can be memorized by the network, although the storage capacity is limited to only two images. Next, it is shown that these restrictions can be uplifted by considering non-reciprocal coupling that allows for treating continuous patterns, while increasing the storage capacity. A simple learning rule for training such coherent laser networks is introduced, which is based on simultaneously embedding a number of patterns as fixed point solutions of the dynamical models governing laser networks. These results are justified by numerical simulation of the dynamical equations governing laser networks.

\section{Formulation}

\subsection{A Single Laser}

Given the importance of a single laser oscillator as an artificial neuron and a building block of the coherent laser network, first we discuss it in the following. Here, laser oscillations is modeled through a second-order nonlinear oscillator as: \cite{mandel1995optical, amnon1989quantum}
\begin{equation}
\label{eq_single}
    \dot{a} = - a + g_0(1-|a|^2)a + b + \xi(t)
\end{equation}
where, $a$ is the complex modal amplitude of the electric field in laser cavity, $g_0$ is the small signal gain, $b$ represents the complex amplitude of a drive laser, and $\xi$ represents fluctuations. Here, the oscillation frequency $\omega_0$ is gauged out for simplicity, the laser is assumed to be frequency-locked with the drive, and the time is normalized to the photon lifetime, $1/\gamma$, where $\gamma$ is the passive cavity decay rate. This model, which is similar to the single-sideband Van der Pol \cite{van1922lxxxv} or the so-called Staurat-Landau oscillator \cite{stuart1958non}, represents a class-A laser, in which the field decay rate is much less than the decay rates of the atomic degrees of freedom, i.e., atomic polarization and population inversion \cite{tredicce1985instabilities}. The analysis presented in this work is based on this minimal model which facilitates integrability. However, it is later discussed that the results are applicable to a more general class of laser systems.

\begin{figure}
\centering
\includegraphics[width=1.0\linewidth]{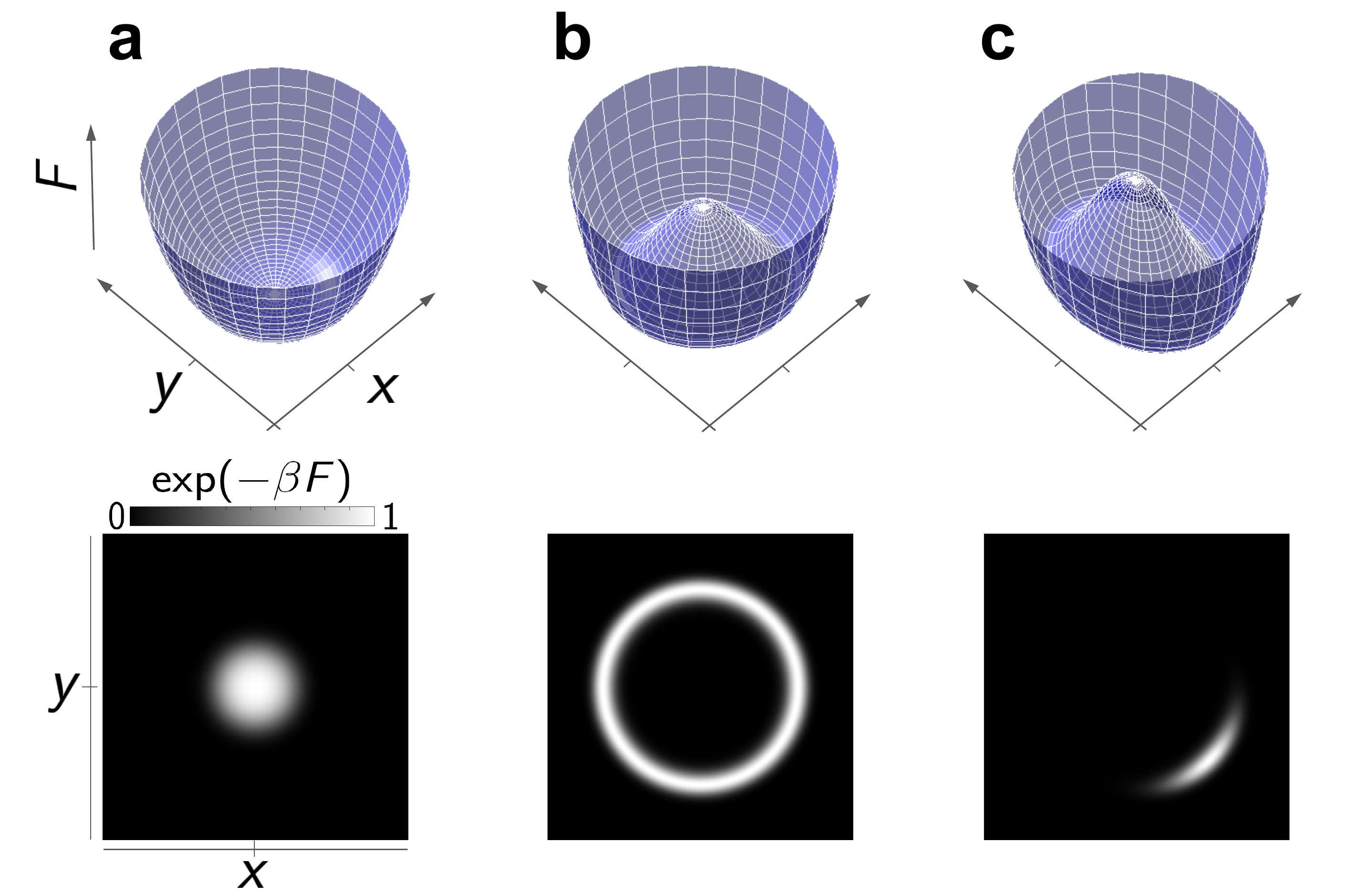}
\caption{The Lyapunov function of a single laser (a) un-pumped, (b) pumped, (c) pumped and seeded, in the in-phase and quadrature phase coordinates, $x=(a+a^*)/\sqrt{2}$ and $y=(a-a^*)/\sqrt{2} i$. (d) The amplitude of the equilibrium solution of a single oscillator versus the pump parameter $g_0$.}
\label{fig2}
\end{figure}

In the absence of seeding, i.e., $b=0$, equation~\ref{eq_single} admits a stable fixed point at $\bar{a}=0$ for $g_0<1$. For $g_0>1$, the stationary solution at zero becomes unstable, while the oscillator stabilizes at $\bar{a}=\sqrt{(g_0-1)/g_0}  \exp(i \phi)$, where, $\phi$ is an arbitrary phase. In the presence of seeding with a complex amplitude $b=|b| \exp(i \varphi)$, the equilibrium state becomes $\bar{a} = |\bar{a}| \exp(i \varphi)$. Therefore, while in the absence of seeding the phase is random, seeding can fix the phase of the laser to that of the drive laser, irrespective of the initial conditions and fluctuations. This aspect is best described in the phase space. By considering a Lyapunov function $F = (g_0-1)|a|^2 - (g_0/2) |a|^4 - (a^* b + a b^*)$, relation~(\ref{eq_single}) is written as $\dot{a} = -\partial F / \partial a^* + \xi (t)$. The governing Lyapunov function is plotted in Fig.~\ref{fig2} for three different scenarios of operating below threshold, above threshold, and in presence of seeding.

\subsection{Laser Networks}

The extension of the dynamical model to the case of $n$ coupled laser oscillators is straightforward. Considering $n$ identical oscillators, the evolution equations can be written as:
\begin{equation}
\label{eq_net}
    \dot{\mathbf{a}} = - \mathbf{a} + g_0 ( 1 - \mathbf{a}^{*} \cdot \mathbf{a}) \cdot \mathbf{a} - W \mathbf{a} + \mathbf{b} + \boldsymbol{\xi} (t).
\end{equation}
In this relation, `$\cdot$' shows entry-wise product, $\mathbf{a} = (a_1, \cdots, a_n)^t$ represents the oscillator amplitudes, $W$ is the coupling matrix, $\mathbf{b} = (b_1, \cdots, b_n)^t$ is the seeding vector, and $\boldsymbol{\xi} (t) = (\xi_1(t), \cdots, \xi_n(t))^t$ contains the fluctuation terms. It is important to note that in a passive coupled cavity arrangement the coupling coefficients are subject to the power conservation and reciprocity relations, which respectively demand $(W - \textup{diag} (W))^{\dagger} (W - \textup{diag} (W)) = 2~ \textup{diag} (W)$ and $W^t = W$. In the following, it is first assumed that the coupling is of pure dissipative nature, thus the matrix elements $w_{ij}$ are taken to be real, and the coupling matrix is assumed to be restricted to the aforementioned conservation relations. A more general case that involves complex coupling coefficients is discussed later.

The symmetry of the coupling matrix allows for writing the dynamical model in terms of the gradient of a Lyapunov function, i.e., $\dot{a}_i = - \partial F / \partial a_i^* + \xi_i(t)$, where \cite{honari2020mapping}
\begin{multline}
\label{eq_Lyap}
    F = - (g_0 -1) \mathbf{a}^{\dagger} \mathbf{a} + \frac{g_0}{2} (\mathbf{a} \cdot \mathbf{a})^{\dagger}(\mathbf{a} \cdot \mathbf{a}) \\ + \frac{1}{2} \mathbf{a}^{\dagger} W \mathbf{a} - (\mathbf{a}^{\dagger} \mathbf{b} + \mathbf{b}^{\dagger} \mathbf{a}),
\end{multline}
It is straightforward to show that along the trajectories of Eq.~(\ref{eq_net}) the time derivative of $F$ is negative. This guarantees that starting from a given set of initial conditions, the evolution of the dynamical system (\ref{eq_net})  is toward the local minima of the multivariate cost function $F(a_1, \cdots, a_n, a_1^*, \cdots, a_n^*)$.

It is important to note that the governing cost function $F$ can be greatly simplified in the strong pump regime, where the amplitudes tend to become uniform and the phase degrees of freedom become the key players in the phase space \cite{honari2020mapping}. This can be seen from Eq.~(\ref{eq_Lyap}), which shows the pump parameter $g_0$ as a penalty for intensity inhomogeneity across the laser network. By directly enforcing the condition of equal equilibrium intensity, i.e., $|a_i| = \sqrt{(g_0-1)/g_0}$, the cost function reduces to the XY Hamiltonian for the phase degrees of freedom:
\begin{equation}
\label{eq_XY}
    f = \sum_{i,j} w_{ij} \cos(\phi_i - \phi_j) - \sum_{i}{|b_i| \cos(\phi_i - \varphi_i)}.
\end{equation}
It is worth recalling that $\phi_i ~ (i=1, \cdots, n)$ represent the phases of the lasers as dynamical variables that describe the phase space of the system, while $\varphi_i ~ (i=1, \cdots, n)$ are constants that represent the phases of the drives. In the following, the attention is focused on the case of the large gain limit, which concerns only the phase degrees of freedom. In addition, for simplicity, the drive term is not considered.

\section{ASSOCIATIVE MEMORY}

The cost function of Eq.~(\ref{eq_XY}) is in general a non-convex function, thus, a coherent laser network with a given weight matrix $W$ could have numerous local minima with different basins of attractions in the phase space. In this case, if the initial point in the phase space is located within the attractor basin of a local minimum, the network will evolve toward the associate stationary state, say $\bar{\Phi} = (\bar{\phi}_1, \cdots, \bar{\phi}_n)^t$. For memorizing a given pattern in the network, the inverse problem is of interest. In this case, the weight matrix $W$ should be devised such that a desired pattern $\Theta = ({\theta}_1, \cdots, {\theta}_n)^t$ becomes a local minimum of the energy function governing the network. In addition, when more than one patterns are to be memorized, of interest is to find a weight matrix $W$ that guarantees the local minima associated with the patterns are located far apart in the phase space such that they can be successfully retrieved. These aspects form the core of training an associative memory, and are discussed in the following.

The cases of binary and continuous patterns are to be treated separately. First, the case of binary pattern, e.g., $\Theta=(\theta_1, \cdots, \theta_n)^t$, where each pixel is limited to two discrete values with contrast $\pi$, say $\theta_i = \pm \pi/2 $, is considered. Next, the analysis is extended to the general case that can treat continuous phase patterns, e.g., $\Theta=(\theta_1, \cdots, \theta_n)^t$, where each pixel takes continuous values, $-\pi \leq \theta_i < +\pi$.

\subsection{Binary Patterns}

As mentioned earlier, the goal of the training is to find the coupling matrix $W$ that results in the presence of local minima of the energy landscape function $f$ (Eq.~\ref{eq_XY}) at desired points. To draw this connection, it is easier to start with identifying the stationary points of the energy landscape function $f$. Enforcing the condition of stationary solutions ${\nabla} f \equiv 0$, results in the following stationary phase relations for the fixed points:
\begin{equation}
\label{eq_stationary_phase}
    \sum_{j} w_{ij} \sin(\bar{\phi}_i - \bar{\phi}_j) = 0 ~~ ; ~~ i=1,\cdots,n
\end{equation}
Clearly, the stationary state condition is satisfied for any binary pattern $\bar{\phi}_i = \theta_i = \pm \pi/2 ~ ; ~ i = 1, \cdots, n$, for any weight matrix. This, however, does not guarantee the presence of stable local minima at such stationary points. On the other hand, a proper weight matrix can be identified that ensures a desired pattern $\Theta$ is a local minimum. This is given by:
\begin{equation}
\label{eq_Heb}
    w_{ij}=-\frac{1}{n} \cos(\theta_i - \theta_j),
\end{equation}
for $i \neq j$. This weight matrix clearly respects the reciprocity condition, i.e., $w_{ij} = w_{ji}$, while the energy conservation can be enforced by choosing the diagonal elements as $w_{ii} = \sum_{j}{|w_{ij}|}$. It can be shown that for the weight matrix given by Eq.~(\ref{eq_Heb}), the desired pattern is a local minimum. This can be shown by using this weight matrix in the XY Hamiltonian of Eq.~(\ref{eq_Heb}), which results in $f = - \sum_{i,j} \frac{1}{n} \cos(\theta_i - \theta_j) \cos(\phi_i - \phi_j)$. Now, one can show that the associated Hessian matrix $H$, with matrix elements $h_{ij} = {\partial}^2 f / \partial \phi_i \phi_j$ at $\phi_i = \theta_i ~ ; ~ i = 1, \cdots, n$ is positive semi-definite, which, in turn ensures that the desired pattern is a stable local minimum of the XY Hamiltonian (see Methods).

\begin{figure}
    \centering
    \includegraphics[width=1.0\linewidth]{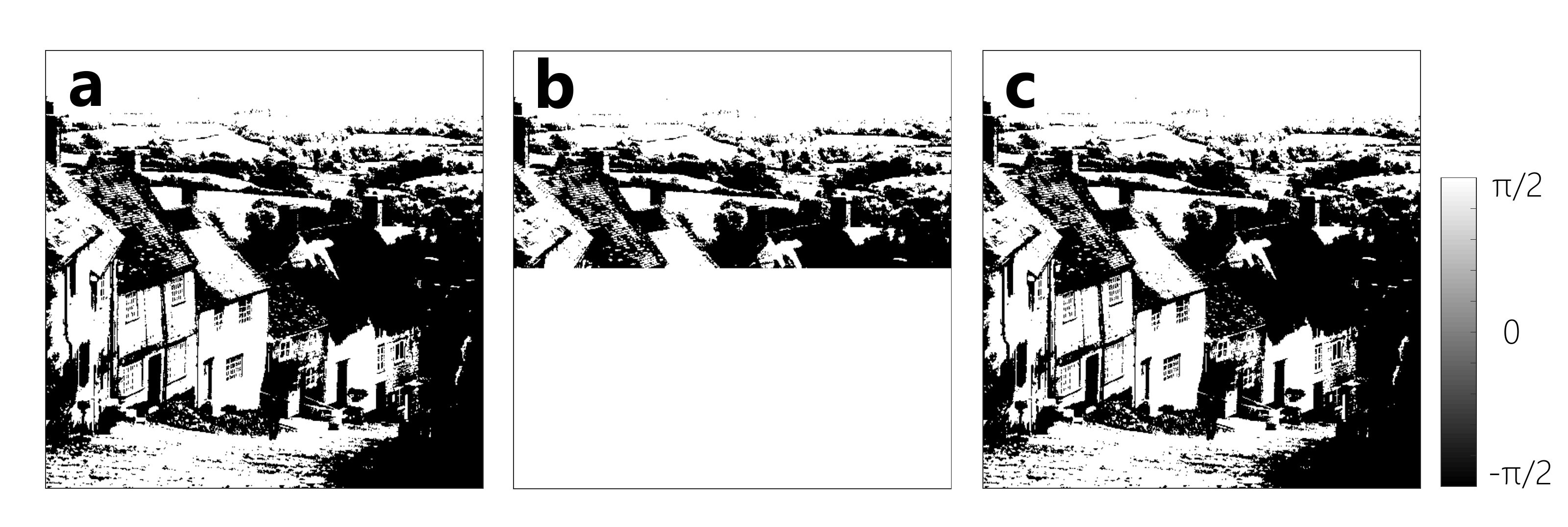}
    \caption{Reconstruction of binary images by CLNs with proper weight matrices. (a) A binary $512 \times 512$ pixel image memorized by the CLN. (b) A portion of the memorized binary image is used as initial phase distribution of the trained CLN. (c) The original image is reconstructed as the CLN evolves to an equilibrium state.}
    \label{fig3}
\end{figure}

Figure~(\ref{fig3}) depicts the reconstruction of a binary pattern in a coherent laser network trained according to Eq.~(\ref{eq_Heb}). Here, a binary $512 \times 512$ pixel image is considered (Fig.~\ref{fig3}(a)). A truncated version of the image is considered as the initial phases of the oscillators (Fig.~\ref{fig3}(b)), and the network successfully retrieves the original image after reaching an equilibrium (Fig.~\ref{fig3}(c)). It is worth noting that in practice, the initial phases might not be controllable, while instead seeding can be used to suitably drive the network toward the memorized pattern. 

It is worth mentioning the similarity of the laser network with the Hopfield network in case of binary patterns. For binary values with $\pi$ contrast, the XY Hamiltonian of Eq.~(\ref{eq_XY}) becomes equivalent with the Ising Hamiltonian $\sum_{i,j} w_{ij} s_i s_j$ ($s_i = \pm 1$), which forms the basis of the Hopfield network. Similarly, the weight matrix given by Eq.~(\ref{eq_Heb}) becomes equivalent to the Hebbian learning rule of the Hopfield network, i.e., $w_{ij} = - \frac{1}{n} s_i s_j$ \cite{hopfield1982neural}. However, this similarity could be misleading given that the phase model discussed above is fundamentally different from the Hopfield network. In fact, in the dynamical model proposed by Hopfield, often called the Hopfield-Tank network, nonlinear activation functions enforce binary operation of the underlying neurons, which instead allows for physical realization of a combinatorial model \cite{hopfield1984neurons, hopfield1985neural}. On the other hand, in the phase model discussed above, the neurons individually operate in continuous phases. In fact, here the formation of a binary pattern is solely a collective behavior that happens as a result of embedding such a pattern as a local minimum of the XY Hamiltonian through a proper design of the weight matrix. This aspect results in a fundamental challenge in using the XY model with real-valued weights as an associative memory as discussed in the following.

The Hebbian learning of Eq.~(\ref{eq_Heb}) can be readily generalized to store more than one pattern. In this case, for $k$ given patterns $\{\Theta^{(1)}, \cdots, \Theta^{(k)}\}$, where, $\Theta^{(l)}=(\theta_1^{(l)},\cdots,\theta_n^{(l)})^t$ are $n$-dimensional binary phase vectors, the weight matrix is chosen as $w_{ij} = - \frac{1}{k} \frac{1}{n} \sum_{k} \cos(\theta_i^{(k)} - \theta_j^{(k)})$. However, it is shown that the learning capacity of such a network of phase oscillators is very limited \cite{cook1989mean, arenas1994phase}. In fact, using a mean-field formalism it is proven that in a network trained with the aforementioned weights, the landscape function $f = - \frac{1}{k} \frac{1}{n} \sum_{k} \sum_{i,j} \cos(\theta_i^{(k)} - \theta_j^{(k)}) \cos(\phi_i - \phi_j)$ exhibits a large number of local minima \cite{cook1989mean}. However, these local minima have significant overlap which prevents successful retrieval of the memorized patterns.


It is worth noting that in case of non-binary patterns the weight matrix of equation~(\ref{eq_Heb}) does not guarantee that a desired continuous pattern is a stationary point. However, it guarantees local convexity of the landscape function at that point (see Methods). Accordingly, a network trained with relation~(\ref{eq_Heb}) can evolve into a nearby local minimum, which, given the highly non-convex nature of the landscape function could be close to the desired pattern. The exact reconstruction of continuous patterns is possible by utilizing complex coupling as discussed next.

\subsection{Continuous Patterns}

The challenge with embedding a continuous pattern as a stable local minimum of the XY Hamiltonian of Eq.~(\ref{eq_XY}) can be resolved by making a simple change in the form of the Hamiltonian as suggested in Ref.~\cite{cook1989mean}. This is done by considering the training parameters as a phase factor in the sinusoidal function according to $f = \sum_{i,j} \cos(\phi_i - \phi_j - \psi_{ij})$, where the network can be simply trained to exhibit a stable local minimum at the desired continuous pattern $\Theta = (\theta_1, \cdots, \theta_n)^t$ by the choice of $\psi_{ij} = \theta_i - \theta_j$. It is important to note that this simple change in the phase cost function demands for complex and non-reciprocal coupling among the lasers that is to be discussed later. In addition, its generalization to storing more than one patterns, according to $f=\frac{1}{k} \frac{1}{n} \sum_{k} \sum_{i,j} \cos[(\phi_i - \phi_j) - (\theta_i^{(k)} - \theta_j^{(k)})]$, suffers from large overlap between the memories \cite{cook1989mean, arenas1994phase}.

Inspired by the clock model proposed in Ref.~\cite{cook1989mean}, here the following modification of the XY Hamitlonian is suggested:
\begin{equation}
\label{eq_XY_Mod}
    f = \sum_{i,j} |w_{ij}| \cos(\phi_i - \phi_j - \psi_{ij} ) - \sum_{i}{|b_i| \cos(\phi_i - \varphi_i)},
\end{equation}
where, $w_{ij} = |w_{ij}| \exp(i \psi_{ij})$ are complex weights. This energy function contains additional parameters, i.e., the amplitudes and phases of the weight matrix elements, which can be trained to store multiple patterns. In the following, it is shown that this phase cost function can be effectively mapped onto a coherent laser network by uplifting the physical limitations of the coupling matrix.

Considering a given continuous pattern $(\theta_1,\cdots,\theta_n)^t$ as the equilibrium phase pattern of a laser network, and assuming that the lasers reach uniform intensities, the associated stationary state complex field amplitude is $\mathbf{\bar{a}} \equiv (e^{i\theta_1}, \cdots, e^{i\theta_n})^t$. To make this a fixed point of the dynamical model governing the coherenet laser network, i.e., $d\mathbf{\bar{a}} /dt \equiv 0$, one needs to ensure $W \mathbf{\bar{a}} = \mathbf{\bar{0}}$. This relation can be solved for $W$, which gives result to $W = C ( I - \mathbf{\bar{a}} \mathbf{\bar{a}}^{+}$), where, $C$ is an arbitrary $n \times n$ matrix, $I$ represents an $n \times n$ identity matrix, and $\mathbf{\bar{a}}^{+}=\mathbf{\bar{a}}^{\dagger}/\mathbf{\bar{a}}^{\dagger}\mathbf{\bar{a}}$ is the pseudoinverse of $\mathbf{\bar{a}}$. For the straightforward choice of $C=I$, the elements of the weight matrix, $w_{ij} = \delta_{ij} - \frac{1}{n} \exp [i(\theta_i - \theta_j)]$, are complex and respecting $w_{ij}=w_{ji}^*$. In this case, apart from the diagonal elements, the elements of the coupling matrix have uniform amplitudes. However, as discussed next, the amplitudes $|w_{ij}|$ play an important role when more than one patterns are involved. Next, consider memorizing $k$ patterns $\{ \Theta^{(1)},\cdots,\Theta^{(k)} \}$, where $\Theta^{(l)}=({\theta_1^{(l)}},\cdots,\theta_n^{(l)})^t$. The desired stationary state complex field vectors are $\mathbf{\bar{a}}^{(l)} = (e^{i\theta_1^{(l)}}, \cdots, e^{i \theta_n^{(l)}})^t$ which can be cast as columns of an $n \times k$ matrix $A = [\mathbf{\bar{a}}^{(1)}, \cdots, \mathbf{\bar{a}}^{(k)}]$. To make these patterns stationary states of the laser network, one needs to enforce the condition of $ W A \equiv \mathbf{0}$, which can be satisfied by the choice of
\begin{equation}
\label{eq_W}
    W = C ( I - A A^{+})
\end{equation}
where, again, $C$ is an arbitrary matrix and $I$ is the identity matrix. A convenient choice is $C=I$ which results in $W =  I - A A^{+}$. 

Assuming that the target $k$ patterns are linearly independent vectors, the weight matrix $W$ is of rank $k$. Therefore, its physical implementation requires $n \times k$ independent matrix elements. In addition, similar to the case of a single pattern, it is straightforward to show that this weight matrix is generally complex but Hermitian, i.e., $W^{\dagger} = W$.

It is important to note that the presence of non-reciprocal coupling ($w_{ij} \neq w_{ji}$) does not generally rule out the possibility of phase locking of the network \cite{fruchart2020phase}. In fact, the Hermiticity of the weight matrix allows the system to admit a Lyapunov function, which guarantees asymptotic stability of the laser network. In this case, due to the Hermiticty of the coupling matrix, $w_{ij}^{*}=w_{ji}$, the Lyapunov function is the same as relation~(\ref{eq_Lyap}). In addition, by taking $a_i = |a_i| \exp(i \phi_i)$ and assuming homogeneous amplitudes, the energy function of relation~(\ref{eq_Lyap}) reduces to the desired phase function of relation~(\ref{eq_XY_Mod}). It is worth stressing that the main difference of the energy function of Eq.~(\ref{eq_XY_Mod}) with the clock Hamiltonian proposed in Ref.~\cite{cook1989mean} is the presence of the amplitudes of the coupling elements $|w_{ij}|$. This additional degree of freedom allows for increasing the storage capacity of the network through the learning rule of Eq.~({\ref{eq_W}}).


\begin{figure*}
    \centering
    \includegraphics[width=0.7\linewidth]{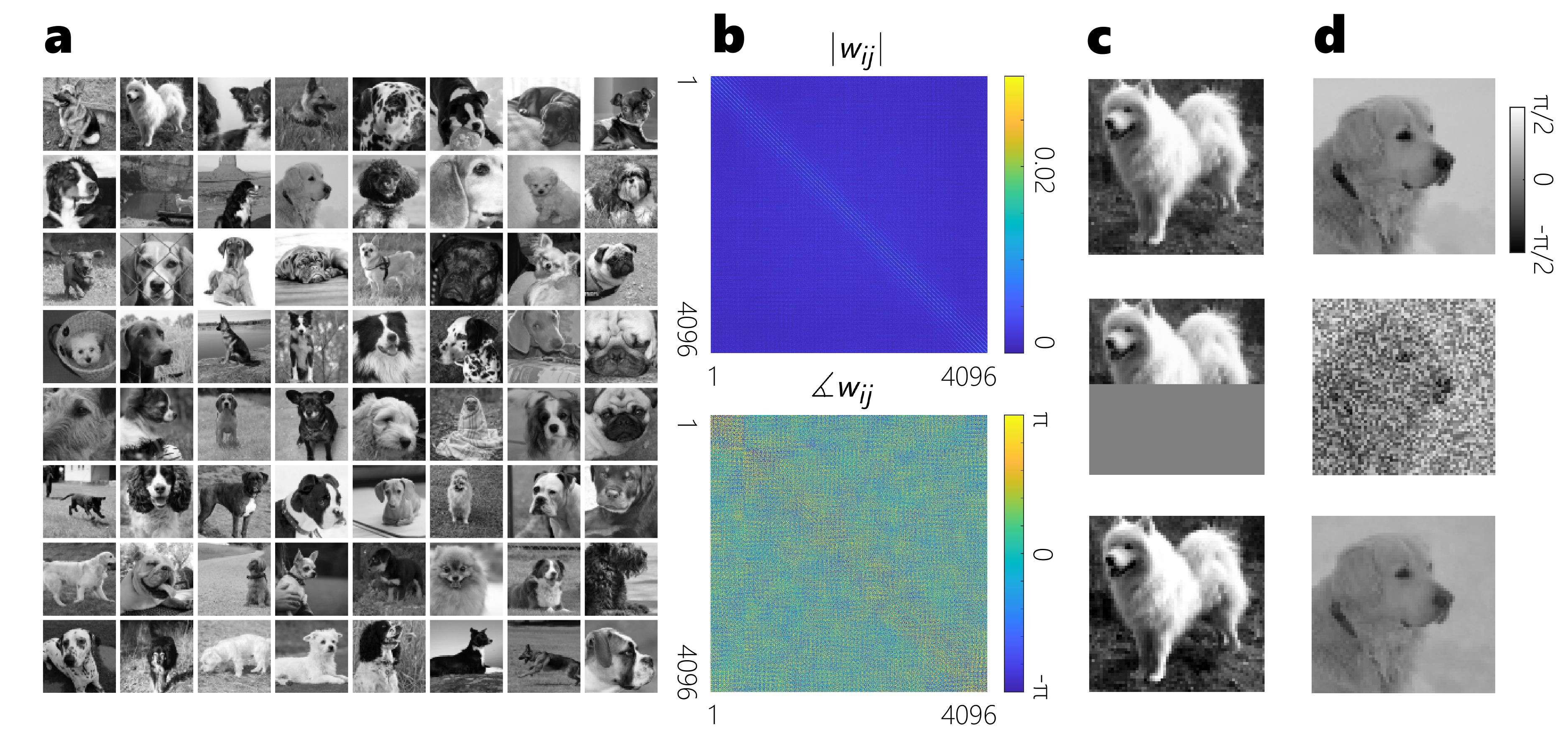}
    \caption{Reconstruction of binary and gray-scale images by CLNs with proper weight matrices. (a) A binary $512 \times 512$ pixel image memorized by the CLN. (b) A portion of the memorized binary image is used as initial phase distribution of the trained CLN. (c) The original image is reconstructed as the CLN dynamically evolves to an equilibrium state. (d-f) Similar to (a-c) for a grayscale $512 \times 512$ pixel image.}
    \label{fig4}
\end{figure*}

The proposed learning is tested with a dataset of $k=64$ continuous patterns of $n = 64 \times 64$ pixels, shown in Fig.~\ref{fig4}(a). These grayscale images are selected from a collection of dog faces from the downsampled ImageNet dataset \cite{ImageNetLink}. The amplitude and phase of the complex weight matrix of Eq.~(\ref{eq_W}) are plotted in Figs.~\ref{fig4}(b). Here, the network successfully stores and retrieves all the $64$ training patterns. For demonstration purposes, the reconstruction of two exemplary images from their corrupted versions is depicted in Figs.~\ref{fig4}(c,d).

\section{Conclusion and Discussion}

In summary, in this paper the potential of using coherent laser networks for neural computing was proposed. The coherent laser network is governed by a non-convex energy landscape function that can contain a large number of fixed point attractors. The use of the coherent laser network as an energy-based neural network model was demonstrated through an associative memory functionality. It was shown that using non-reciprocal coupling between lasers allows for going beyond binary data and adding the capability of handling continuous patterns. This work, outlines the great potential of coherent laser networks for optical neural computing. In addition, the proposed dynamical model could have applications as a novel continuous-time neural network for conventional digital computing.

It is worth mentioning that the results presented above were built on the idealistic assumption of identical oscillators, while in practice, individual laser cavities can have deviations in their resonance frequencies and linewidths. However, simulation results show that the system exhibits self-organizing behavior and can reach phase-locking in presence of tolerable perturbations. To explore this aspect, the network of Fig.~\ref{fig4} is simulated under the presence of random frequency and linewidth detunings of the individual network elements. It is found that the laser network reaches phase-locking and the associative memory functionality is preserved. The results are shown in the supporting Supplementary Material \cite{suppl}.

The results presented in this work were based on the the so-called class-A laser model, where the gain can be considered a constant, while many practical lasers fall in the category of class-B lasers, where the gain evolves dynamically \cite{tredicce1985instabilities}. The simplified model used here admits a Lyapunov function, which allows for an analytical treatment of the laser network and finding a training method. However, it should be noted that the proposed training method concerns solely the stationary behavior of the network through the coupling matrix. Therefore, the dynamics of the gain is not expected to violate the associative memory functionality, so long as the stability of the fixed points are guaranteed. As shown recently, a large gain lifetime, compared to the photon lifetime, can give rise to destabilization of shallow local minima or metastable states such as vortex states in a lattice of coupled lasers \cite{honari2021self}. In this case, however, numerical simulations indicate that the patterns embedded through the learning rule of Eq.~{\ref{eq_W}} remain stable even for large gain lifetimes. This is justified by repeating the simulations of Fig.~\ref{fig4} with a class-B laser model as described in the Supplementary Material \cite{suppl}.

Finally, the present work is focused on the associative memory functionality as a generic task for energy-based models, while it remains to examine the full capacity of coherent laser networks as energy-based models in different network architectures and for various functionalities \cite{goodfellow2016deep}. In addition, the proposed pseudoinverse learning is a simplistic approach, which is suitable for experimental realization given that it requires a low-rank weight matrix. However, of interest would be to develop advanced training algorithms that allow for harnessing the full capacity of the coherent laser networks for machine learning.

\section{METHODS}

\subsection{Numerical Simulations}

The coherent laser as described by equations~(\ref{eq_net}) is in essence a continuous-time energy-based recurrent neural network. Considering the potential importance of the proposed model for unconventional computing through simulations of the underlying model with digital computers, in the following, numerical simulations are briefly discussed. The numerical simulations of Eqs.~(\ref{eq_net}) are performed with a forward-difference Euler method according to:
\begin{equation}
\label{eq_num_complex}
        \mathbf{a}(t+\Delta t) = \Delta t \left[- \mathbf{a}(t) + \mathbf{f}(\mathbf{a}(t)) - W \mathbf{a}(t) + \mathbf{b} + \boldsymbol{\xi} (t) \right].
\end{equation}
For the simulations discussed in this paper, the network converges rapidly (after $\sim 100$ steps). In general, the most computationally-costly process in Eq.~(\ref{eq_num_complex}) is the matrix-vector multiplication. It should also be noted that Eq.~(\ref{eq_num_complex}) deal with complex numbers, which require double-precision floating point format. To consider noise, uncorrelated delta noise is generated for each oscillator, i.e., $\langle \xi_i(t')^* \xi_j(t) \rangle = D \delta_{ij} \delta(t-t')$. The effect of detuing is considering by changing the first term of Eq.~(\ref{eq_num_complex}) according to $ - \mathbf{a}(t) \rightarrow - (\mathbf{1} + \boldsymbol{\delta \gamma} + i \boldsymbol{\delta \omega} ) \cdot \mathbf{a}(t)$, where, $\boldsymbol{\delta \omega} = (\delta \omega_1, \cdots, \delta \omega_n)^t$, and $\boldsymbol{\delta \gamma} = (\delta \gamma_1, \cdots, \delta \gamma_n)^t$. Here, $\delta \omega_i , \delta \gamma_i \sim \mathcal{N}(0,\sigma)$ with $\sigma \sim 0.05$.

\subsection{The Hessian Matrix}

The Lyapunov function of Eq.~(\ref{eq_Lyap}) is a function of $2n$ variables, which can be cast in a vector as $\mathbf{e} = (a_1, \cdots , a_n, a_1^*, \cdots, a_n^*)^t$. The Lyapunov function near an arbitrary point can be expanded as:
\begin{equation}
    F (\bar{\mathbf{e}} + \Delta \mathbf{e}) = F(\bar{\mathbf{e}}) + \Delta \mathbf{e}^t ~ \nabla F(\bar{\mathbf{e}}) + \frac{1}{2} \Delta \mathbf{e}^t ~ H(\bar{\mathbf{e}}) ~ \Delta \mathbf{e} + \cdots
\end{equation}
where, $\nabla F$ is the gradient vector and $H$ is a $2n \times 2n$ Hessian matrix. In this representation, stationary states are points associated with $\nabla F(\bar{\mathbf{e}}) = \mathbf{0}$. The Hessian matrix can be represented as:
\begin{equation}
H=
\begin{pmatrix}
H_d   & H_o \\
H_o^* & H_d^* 
\end{pmatrix}
\end{equation}
where, $H_d = (g_0 - 1) I - 2 g_0 \textup{diag} ( \mathbf{a}^* \cdot \mathbf{a} ) - W$, and $H_o = - g_0 \textup{diag} ( \mathbf{a} \cdot \mathbf{a} )$.

For the phase cost function of Eq.~(\ref{eq_XY}), the Hessian matrix is an $n \times n$ matrix with elements $h_{ij}=\partial^2 f / \partial \phi_i \partial \phi_j$, which are found to be:
\begin{equation}
\label{eq2}
    h_{ij}=\left\{\begin{matrix}
~~~~~~~~~ w_{ij} \cos(\phi_i - \phi_j); ~~ i \neq j
\\ 
- \sum_{j}{} w_{ij} \cos(\phi_i - \phi_j); ~~ i = j
\end{matrix}\right.
\end{equation}
For the choice of the weight matrix of Eq.~(\ref{eq_Heb}) for a given pattern $\Theta = ({\theta}_1, \cdots, {\theta}_n)^t$, evaluating the Hessian at this pattern, results in the off-diagonal elements $h_{ij} = - \frac{1}{n} \cos^2(\theta_i - \theta_j) $, and diagonal elements $h_{ii} = \frac{1}{n} \sum_{j}{} \cos^2(\theta_i - \theta_j)$. In this case, the Hessian matrix is of the form of the Laplacian matrix of a weighted graph with adjacency matrix elements $\frac{1}{n} \cos^2(\theta_i - \theta_j)$ \cite{chung1997spectral}. It is straightforward to show that this Hessian matrix is positive semidefinite given that it is symmetric and diagonally dominant \cite{chung1997spectral}. This result is valid for both choices of binary and continuous patterns, however, one should recall that the training of Eq.~(\ref{eq_Heb}) is limited to patterns that pass the stationary test of Eq.~(\ref{eq_stationary_phase}), that is limited to binary patterns.


%
%


%

\section{Data availability}

The data that support the findings of this study are available from the corresponding author upon reasonable request.

\bibliography{CLN}

\begin{thebibliography}{10}
\newcommand{\enquote}[1]{``#1''}

\bibitem{calude2017unconventional}
C.~S. Calude, \enquote{Unconventional computing: A brief subjective history,}
  in \emph{Advances in Unconventional Computing,}  (Springer, 2017), pp.
  855--864.

\bibitem{haken2004synergetic}
H.~Haken, \emph{Synergetic computers and cognition: A top-down approach to
  neural nets}, vol.~50 (Springer Science \& Business Media, 2004).

\bibitem{hertz2018introduction}
J.~Hertz, A.~Krogh, and R.~G. Palmer, \emph{Introduction to the theory of
  neural computation} (CRC Press, 2018).

\bibitem{hopfield1982neural}
J.~J. Hopfield, {\protect\JournalTitle{Proc. Natl. Acad. Sci. U.S.A.}}
  \textbf{79}, 2554 (1982).

\bibitem{hopfield1984neurons}
J.~J. Hopfield, {\protect\JournalTitle{Proceedings of the national academy of
  sciences}} \textbf{81}, 3088 (1984).

\bibitem{farhat1985optical}
N.~H. Farhat, D.~Psaltis, A.~Prata, and E.~Paek, {\protect\JournalTitle{Applied
  optics}} \textbf{24}, 1469 (1985).

\bibitem{graf1987cmos}
H.~Graf and P.~de~Vegvar, \enquote{A cmos associative memory chip based on
  neural networks,} in \emph{1987 IEEE International Solid-State Circuits
  Conference. Digest of Technical Papers,} , vol.~30 (IEEE, 1987), pp.
  304--305.

\bibitem{verleysen1989analog}
M.~Verleysen and P.~G. Jespers, {\protect\JournalTitle{IEEE Micro}} \textbf{9},
  46 (1989).

\bibitem{hopfield1985neural}
J.~J. Hopfield and D.~W. Tank, {\protect\JournalTitle{Biological cybernetics}}
  \textbf{52}, 141 (1985).

\bibitem{ackley1985learning}
D.~H. Ackley, G.~E. Hinton, and T.~J. Sejnowski,
  {\protect\JournalTitle{Cognitive science}} \textbf{9}, 147 (1985).

\bibitem{marandi2014network}
A.~Marandi, Z.~Wang, K.~Takata, R.~L. Byer, and Y.~Yamamoto,
  {\protect\JournalTitle{Nature Photonics}} \textbf{8}, 937 (2014).

\bibitem{mcmahon2016fully}
P.~L. McMahon, A.~Marandi, Y.~Haribara, R.~Hamerly, C.~Langrock, S.~Tamate,
  T.~Inagaki, H.~Takesue, S.~Utsunomiya, K.~Aihara, R.~L. Byer, M.~M. Fejer,
  H.~Mabuchi, and Y.~Yamamoto, {\protect\JournalTitle{Science}} \textbf{354},
  614 (2016).

\bibitem{inagaki2016coherent}
T.~Inagaki, Y.~Haribara, K.~Igarashi, T.~Sonobe, S.~Tamate, T.~Honjo,
  A.~Marandi, P.~L. McMahon, T.~Umeki, K.~Enbutsu, O.~Tadanaga, H.~Takenouchi,
  K.~Aihara, K.-i. Kawarabayashi, K.~Inoue, S.~Utsunomiya, and H.~Takesue,
  {\protect\JournalTitle{Science}} \textbf{354}, 603 (2016).

\bibitem{nixon2013observing}
M.~Nixon, E.~Ronen, A.~A. Friesem, and N.~Davidson,
  {\protect\JournalTitle{Physical review letters}} \textbf{110}, 184102 (2013).

\bibitem{berloff2017realizing}
N.~G. Berloff, M.~Silva, K.~Kalinin, A.~Askitopoulos, J.~D. Töpfer,
  P.~Cilibrizzi, W.~Langbein, and P.~G. Lagoudakis,
  {\protect\JournalTitle{Nature Materials}} \textbf{16}, 1120 (2017).

\bibitem{parto2020realizing}
M.~Parto, W.~Hayenga, A.~Marandi, D.~N. Christodoulides, and M.~Khajavikhan,
  {\protect\JournalTitle{Nature Materials}}  (2020).

\bibitem{tradonsky2019rapid}
C.~Tradonsky, I.~Gershenzon, V.~Pal, R.~Chriki, A.~Friesem, O.~Raz, and
  N.~Davidson, {\protect\JournalTitle{Science advances}} \textbf{5}, eaax4530
  (2019).

\bibitem{kalinin2018global}
K.~P. Kalinin and N.~G. Berloff, {\protect\JournalTitle{Scientific reports}}
  \textbf{8}, 1 (2018).

\bibitem{honari2020mapping}
M.~Honari-Latifpour and M.-A. Miri, {\protect\JournalTitle{Physical Review
  Research}} \textbf{2}, 043335 (2020).

\bibitem{honari2021self}
M.~Honari-Latifpour, J.~Ding, S.~Takei, and M.-A. Miri,
  {\protect\JournalTitle{arXiv preprint arXiv:2102.10157}}  (2021).

\bibitem{brunner2015reconfigurable}
D.~Brunner and I.~Fischer, {\protect\JournalTitle{Optics letters}} \textbf{40},
  3854 (2015).

\bibitem{mandel1995optical}
L.~Mandel and E.~Wolf, \emph{Optical coherence and quantum optics} (Cambridge
  University Press, 1995).

\bibitem{amnon1989quantum}
A.~Yariv, \emph{Quantum electronics} (Wiley, 1989).

\bibitem{van1922lxxxv}
B.~van~der Pol, {\protect\JournalTitle{The London, Edinburgh, and Dublin
  Philosophical Magazine and Journal of Science}} \textbf{43}, 700 (1922).

\bibitem{stuart1958non}
J.~T. Stuart, {\protect\JournalTitle{Journal of Fluid Mechanics}} \textbf{4}, 1
  (1958).

\bibitem{tredicce1985instabilities}
J.~R. Tredicce, F.~T. Arecchi, G.~L. Lippi, and G.~P. Puccioni,
  {\protect\JournalTitle{JOSA B}} \textbf{2}, 173 (1985).

\bibitem{cook1989mean}
J.~Cook, {\protect\JournalTitle{Journal of Physics A: Mathematical and
  General}} \textbf{22}, 2057 (1989).

\bibitem{arenas1994phase}
A.~Arenas and C.~P. Vicente, {\protect\JournalTitle{EPL (Europhysics Letters)}}
  \textbf{26}, 79 (1994).

\bibitem{fruchart2020phase}
M.~Fruchart, R.~Hanai, P.~B. Littlewood, and V.~Vitelli,
  {\protect\JournalTitle{arXiv preprint arXiv:2003.13176}} \textbf{19} (2020).

\bibitem{ImageNetLink}
\url{http://www.image-net.org/}.

\bibitem{suppl}
\enquote{See supplemental material,} .

\bibitem{goodfellow2016deep}
I.~Goodfellow, Y.~Bengio, A.~Courville, and Y.~Bengio, \emph{Deep learning},
  vol.~1 (MIT press Cambridge, 2016).

\bibitem{chung1997spectral}
F.~R. Chung and F.~C. Graham, \emph{Spectral graph theory}, 92 (American
  Mathematical Soc., 1997).

\end{thebibliography}



\end{document}